# De Sitter Relativity and Cosmological Principle

Leonardo Chiatti[*]

*Medical Physics Laboratory, AUSL VT, Viterbo, Italy 01100*

**Abstract**

The formalism of Fantappié-Arcidiacono Projective General Relativity –also known as De Sitter Relativity- has recently been revised in order to make possible cosmological models with expansion, similarly to ordinary Friedmann cosmology formulated within the context of General Relativity. In this article, several consequences of interest in the current cosmological debate are examined and discussed in a semiquantitative manner. Specifically: re-examination of the Supernova Project results using this new formalism, with a new estimate of the cosmological parameters; the ordinary matter to dark matter densities ratio; the existence of a new fundamental constant having the dimensions of an acceleration and its relation with dark matter.

## 1. INTRODUCTION

Friedmann's cosmological models derive from the application of the cosmological principle to Einstein's General Relativity (GR). The cosmological principle has recently been applied to an appropriately revised form of Fantappié-Arcidiacono Projective General Relativity (PGR) [1], also known as De Sitter Relativity. While reference is made to the original works for the theoretical details [2, 3], here we confine ourselves to summarizing briefly the main differences between these two theories. PGR introduces a new fundamental constant of nature which does not exist in GR; dimensionally, this new constant is a time interval and will be indicated here with $t_0$. PGR is a generalization of GR and, for $t_0 \to \infty$, collapses into ordinary GR. The limit of the two theories in the case of vanishing density of matter is constituted by Special Relativity (SR) for GR without a cosmological term and by Projective Special Relativity (PSR) for PGR, respectively. PSR leads to a De Sitter chronotope having radius $r = ct_0$ with $c$, as is customary, being the speed of light in a vacuum [4, 5, 6, 7].

This chronotope can be conceived as being derived from the Arcidiacono 5-sphere, described in projective coordinates by the equation:

$$(\underline{x}_0)^2 + (\underline{x}_1)^2 + (\underline{x}_2)^2 + (\underline{x}_3)^2 + (\underline{x}_5)^2 = r^2 \quad , \tag{1}$$

through a Wick rotation[1] of the time coordinate $\underline{x}_0$:

$$(\underline{x}_0)^2 - (\underline{x}_1)^2 - (\underline{x}_2)^2 - (\underline{x}_3)^2 + (\underline{x}_5)^2 = r^2 \quad . \tag{2}$$

The space expansion is described by the following canonical extension of Eq. (2):

$$(\underline{x}_0)^2 - R^2(\tau)[(\underline{x}_1)^2 + (\underline{x}_2)^2 + (\underline{x}_3)^2] + (\underline{x}_5)^2 = r^2 \quad , \tag{3}$$

---

[*] Address correspondence to this author at AUSL Viterbo, Medical Physics Laboratory, Via Enrico Fermi 15, 01100 Viterbo (Italy). E-mail: fisica1.san@asl.vt.it

[1] Here followed by an inessential inversion of the signature of spacetime coordinates alone.

which leads back to Eq. (2) when the trivial scale distance $R(\tau) \equiv 1$ [$\tau$ being the cosmic time] is adopted. Eq. (3) can be coupled with the PGR gravitational equations, formulated by Arcidiacono as far back as 1964 [8], for the purpose of describing a homogeneous cosmic fluid in a space undergoing expansion. The equivalent is thus obtained - in the context of PGR - of that which Friedmann cosmology represents for GR.

An important difference with respect to Friedmann cosmology is that whilst the latter admits of a multiplicity of possible models, to be subsequently selected based on observation, the approach described here seems to lead to a single cosmological model [3]. It corresponds to the Friedmann model having null spatial curvature ($k = 0$) and a positive cosmological term $\lambda = 4/3t_0^2$.

The reduction of arbitrariness is not however the only interesting aspect of this approach. At the start of the expansion [$R(\tau) = 0$], Eq. (3) becomes:

$$(\underline{x}_0)^2 + (\underline{x}_5)^2 = r^2 \quad ; \tag{4}$$

if it is therefore assumed that the start of the expansion coincides with the origin of $\underline{x}_0$, i.e. that the big bang occurs on the equator $\underline{x}_0 = 0$ of the hypersphere (1), the value $\pm r$ is obtained for the variable $\underline{x}_5$. In geometrical terms, this corresponds to a pointlike big bang associated with a point on the equator of the Arcidiacono 5-sphere. However, the $\underline{x}_5$ axis can be rotated on this equator giving rise to $\infty^3$ different (and equivalent) intersections. One thus has $\infty^3$ different (and equivalent) big bangs or, to be more precise, $\infty^3$ different (and equivalent) views of the same big bang, which are pertinent to distinct fundamental (inertial) observers.

In an individual observer's coordinates the metric is consistent with Eq. (3) and therefore all the observers see a Universe in expansion. At a certain value of cosmic time $\tau$, all the observers see the Universe under the same conditions and the cosmological principle thus applies, provided that the conditions of matter on the equator $\underline{x}_0 = 0$ are homogeneous.

The dimensionless vacuum starting from which the big bang develops is therefore substituted, in this approach, by a pre-existing space: the equator of the 5-sphere (1). The passage from condition (1) to condition (3) takes place at a critical value $\theta_0$ of the variable $\underline{x}_0/c$ for which processes of quantum localization of elementary particles on spacetime become possible. The sudden appearance of all the elementary particles on the spacetime domain is the true essence of the big bang. Starting from this nucleation, the propagation of particles is described by wavefunctions in which coordinates satisfying condition (3) and no longer condition (1) appear as an argument. The "archaic phase" governed by condition (1) comes to an end and the actual history of the Universe governed by condition (3) begins.

It is plausible that $c\theta_0$ coincides with the classical electron radius [3], which also corresponds to the range of strong interactions. If this identification is correct, the nucleation of matter originates a spatially extended big bang, corresponding to the condition under which, in the *conventional* big bang model, the Universe would find itself at the end of the hadronic era.

We can therefore recapitulate the idea as follows. A public "archaic" spacetime exists, whose coordinates satisfy Eq. (1). The distribution of matter is homogeneous with respect to the "public" coordinates $\underline{x}_1, \underline{x}_2, \underline{x}_3$ and depends solely on the coordinate $\underline{x}_0$; matter is in a state of virtual quantum processes and this homogeneity is thus nothing other than the homogeneity of a sort of pre-cosmic vacuum. The quantum fluctuations originating at the equator $\underline{x}_0 = 0$ and having a size of $\underline{x}_0 = c\theta_0$ end with quantum localization processes of particles on a chronotope which each observer represents through his own "private" coordinates, normalized in accordance with Eq. (3). From an individual observer's viewpoint and in conformity with the private coordinates adopted by him, the phase transition appears as a spatially extended big bang of finite density. Starting from this "fire sphere" (which, if $c\theta_0$ is interpreted as the classical radius of the electron, is the same as in the conventional theory considered at the end of the hadronic era) an expansion of "private" space develops that is consistent with a flat Friedmann model having a positive cosmological constant. This applies for all fundamental observers and the cosmological principle is complied with. The initial homogeneity is expressed in a public spacetime as a non-dynamically generated initial condition; it is reflected in the private spacetimes of individual fundamental observers.

This, very briefly, is the proposed scenario. In the following sections various issues connected with the adoption of this scenario are examined; the attempt will be made to highlight the contribution it can give to the contemporary cosmological debate.

## 2. THE PRIMEVAL FIRE SPHERE

Let us denote with $E$ the energy released during the nucleation process; "before" nucleation occurs (i.e. for $\underline{x}_0 < c\theta_0$) this energy is distributed in latent form on the hypersphere equator. If we indicate the volume of the equator with $V = 2\pi^2 r^3$, the average density of energy is expressed, in the "public" metric, by $E/V$ and is finite. In the "public" coordinate $\underline{x}_0$ nucleation starts at $\underline{x}_0 = c\theta_0$ and has a finite extension $\approx c\theta_0$; it occurs at an exponential rate of $(\theta_0)^{-1}$ and not instantaneously. In conclusion, in the description offered by public metric the nucleation process is not associated with discontinuity of dynamics or with infinite densities and thus there are no singularities.

In the description offered by the private metric of a specific observer the radius of the space at the end of nucleation is $cf\theta_0$, where $f$ is a constant. The energy density in this second description is $E/(cf\theta_0)^3$ and is still finite; this description, too, is therefore free of singularities.

The two descriptions are connected. In the "private" description the spatial coordinates are contracted because of the scale factor $(cf\theta_0)/r$; thus the density at the end of nucleation is equal to the product of $(E/V)$ by $[(cf\theta_0)/r]^{-3}$, and this corresponds to the result given above. The coefficient $f$ defines here the radius of the Universe at the end of nucleation.

If $c\theta_0$ is interpreted as the classical radius of the electron, then the radius $cf\theta_0$ corresponds to that assigned by the standard big bang model to the fire sphere at the end of the hadronic era. It is to be noted that the baryons/antibaryons ratio must have been already defined at this instant, because subsequent interconversion of the two components is not possible, since the temperature has become too low. Thus the baryonic dominance emerges at nucleation and any mechanisms which determine it - if it is not to be considered an initial condition – act in the "previous" $\underline{x}_0 < c\theta_0$ phase.

## 3. RELATION WITH THE "COSMIC CONCORDANCE" HYPOTHESIS

The Friedmann model $k = 0$, $\lambda > 0$ obtained from PGR is the same as that invoked by the "cosmic concordance" hypothesis. The additional relation $\lambda = 4/3t_0^2$ does not appear to lead to any conflict with the framework commonly accepted today, as shown in the following numeric example.

Starting from the equations for a dust-dominated Universe (with the customary meaning of the symbols):

$$\left(\frac{\dot{R}}{R}\right)^2 = H^2 = \frac{8\pi G \mu_0}{3R^3} + \frac{\lambda}{3} ,$$

$$R^3 = \frac{4\pi G \mu_0}{\lambda c^2}\left[\cosh(\tau'\sqrt{3\lambda}) - 1\right] ,$$

where $\mu_0$ is the density of matter when $R = 1$, one has – by substituting the second equation into the first:

$$H^2 = \frac{2}{3}\frac{\lambda}{\cosh(\tau'\sqrt{3\lambda})-1} + \frac{\lambda}{3}.$$

If one supposes that the current value of regraduated cosmic time (see next section) is $\tau' = 1.29 t_0$, one obtains a value of $H$ at present time equal to:

$$H_0^2 = \frac{2}{3}\frac{\lambda}{\cosh(1.29 t_0 \sqrt{3\frac{4}{3t_0^2}})-1} + \frac{\lambda}{3} = \lambda\left(\frac{2}{3[\cosh(2.58)-1]} + \frac{1}{3}\right) = 0.452\lambda.$$

Assuming thus the accepted value $\lambda = 18 \times 10^{-36}$ h$^2$ s$^{-2}$ one obtains $H_0 = 0.3 \times 10^{-17}$ h s$^{-1}$, which is reasonable. Furthermore:

$$\Omega_\Lambda = \frac{\lambda}{3H_0^2} = \frac{1}{3 \times 0.452} = 0.738.$$

By rewriting the first equation in the form $\Omega = \Omega_\Lambda + \Omega_M = 1$, one thus has $\Omega_M = 1 - 0.738 = 0.262$, which agrees very well with the "concordance" model.
One finally has $t_0 = (4/3\lambda)^{1/2} = 0.27 \times 10^{18}$ h$^{-1}$ s, which is also a reasonable estimate. Actually, as will be shown in the next section, the values of the cosmological parameters must undergo slight modification if the PGR viewpoint is adopted.

## 4. SPECIFYING THE CONCORDANCE MODEL: A NEW ANALYSIS OF DATA FROM SUPERNOVAE Ia

### 4.1 Preliminary remarks

A key difference between the PGR-based approach and that based on ordinary GR lies in the notion of cosmic time. In PGR two cosmic times $\tau$ and $\tau'$ exist, as in Milne's Kinematic Relativity. In accordance with the terminology introduced by Milne [9], now of mere historical importance, $\tau$ and $\tau'$ are respectively called "kinematic time" and "dynamic" or "atomic time". More specifically, in the limit case of an empty Universe $\tau$ and $\tau'$ are connected by the same logarithmic relation as that existing between kinematic time and atomic time in Milne's theory [1, 3]. In general, however, the relation is more complex [3, 10] and we propose to describe it here.
Let us first of all see the meaning of these two times. If the cosmic clock of fundamental observers is graduated according to kinematic time $\tau$, and they use this time to coordinate events, then they detect the recession motion of galaxies associated with spacetime curvature along the direction of time (the spatial curvature is null). In other words, they detect the expansion resulting from the existence of a De Sitter horizon, generated by the finite value of $t_0$. If, on the other hand, fundamental observers use a cosmic clock graduated in accordance with atomic time $\tau'$, then the purely kinematic De Sitter expansion disappears.
Thus, in kinematic time the cosmic recession depends partly on the expansion of space (variation of the scale function) and partly on the finite value of $t_0$. Instead, in atomic time, the cosmic recession depends solely on the expansion of space. The cosmic time usually used in Friedmann models is atomic time.
The regraduation of the cosmic clock of fundamental observers from kinematic time to atomic time gives rise to the appearance of the cosmic repulsion associated with the cosmological term. This is the origin of the cosmological term in this framework [3, 6, 10].

## 4.2 Relation between kinematic time and atomic time

Let us indicate with a prime the derivation with respect to $\tau$ and with a dot the derivation with respect to $\tau'$. The first Friedmann equation can thus be written:

$$\frac{8\pi G\mu}{3} = \left(\frac{R'}{R}\right)^2 = \left(\frac{\dot{R}}{R}\right)^2 - \frac{1}{3}\lambda \qquad (5)$$

Multiplying by $R^2$ one obtains:

$$\left(\frac{dR}{d\tau}\right)^2 = \left(\frac{dR}{d\tau'}\right)^2 - \frac{R^2\lambda}{3} = \left(\frac{dR}{d\tau}\right)^2\left[\left(\frac{d\tau}{d\tau'}\right)^2 - \frac{\lambda}{3}\left(\frac{R}{R'}\right)^2\right]$$

i.e.:

$$\left(\frac{d\tau}{d\tau'}\right)^2 = 1 + \frac{\lambda}{3}\left(\frac{R}{R'}\right)^2 \quad .$$

Bearing in mind that $R = \text{constant} \times \tau^{2/3}$, one has $(R/R') = 3\tau/2$ and:

$$\frac{d\tau}{d\tau'} = \sqrt{1 + \frac{3}{4}\lambda\tau^2} \quad . \qquad (6)$$

Eq. (6) defines the relation existing between the intervals of atomic time and kinematic time at the kinematic time instant $\tau$, and can be inverted so as to obtain:

$$\frac{d\tau'}{d\tau} = \frac{1}{\sqrt{1 + \frac{3}{4}\lambda\tau^2}} \quad . \qquad (7)$$

Eq. (5) can be written in the form:

$$\frac{8\pi G\mu}{3} = \left(\frac{R'}{R}\right)^2 = \frac{4}{9}\tau^{-2} \quad \Rightarrow \quad \mu = \frac{\tau^{-2}}{6\pi G} \propto R^{-3} \quad . \qquad (8)$$

Let us consider a ray of light emitted by a galaxy at the kinematic time instant $\tau_{em}$, received by an observer at the kinematic time instant $\tau_{arr}$. The spatial distance travelled by the ray is expressed by the integral:

$$l = \int_{\tau_{em}}^{\tau_{arr}} \frac{c\,d\tau}{R(\tau)} = \int_{\tau_{em}}^{\tau_{arr}} \frac{c\,d\tau}{(\tau/\tau_{arr})^{2/3}} = 3c\tau_{arr}\left[1-\left(\frac{\tau_{em}}{\tau_{arr}}\right)^{1/3}\right]. \qquad (9)$$

In integral (9) it has been assumed that the scale function, evaluated at the time of reception of the signal, is equal to unity.

### 4.3 Redshift

Let us suppose having a spectral line whose frequency is $v$ in kinematic time $\tau$, corresponding to the frequency $v'$ in the atomic time $\tau'$. The number of oscillations $dn$ corresponding to this spectral line will be expressed by the relation $dn = v\,d\tau = v'\,d\tau'$. The ratio $v/v'$ is then expressed by Eq. (7).
Redshift is measured as the percentage difference between the frequency of the line in the radiation coming from a remote galaxy and the frequency of the same line in a spectrum obtained from a local sample; this clearly includes both the spatial scale stretching effect and the projective effect resulting from time curvature. Redshift is therefore expressed through the $v$ frequencies.
However, the ratio of scale distance values upon arrival and at emission should be correlated with redshift expressed as a function of the "atomic" frequency $v'$, because only in atomic time redshift depends entirely on spatial stretching.
Thus:

$$\frac{R_{arr}}{R_{em}} = 1 + \frac{v'_{em} - v'_{arr}}{v'_{arr}} = 1 + \frac{v_{em}\left(\frac{d\tau}{d\tau'}\right)_{em} - v_{arr}\left(\frac{d\tau}{d\tau'}\right)_{arr}}{v_{arr}\left(\frac{d\tau}{d\tau'}\right)_{arr}} = \frac{v_{em}}{v_{arr}}\left(\frac{d\tau'}{d\tau}\right)_{em}^{-1}\left(\frac{d\tau'}{d\tau}\right)_{arr}.$$

i.e.:

$$\frac{\left(R\frac{d\tau}{d\tau'}\right)_{arr}}{\left(R\frac{d\tau}{d\tau'}\right)_{em}} = (1+z)_{uncorrected}. \qquad (10)$$

If one has available an estimate of the distances of a certain number of extragalactic objects whose redshift is known, the following algorithm can be applied.

1) Assuming certain values of $\lambda$ and $\tau_{arr}$, from the distance $l$ the emission time $\tau_{em}$ is obtained by inverting Eq. (9);

2) by substituting $\tau_{arr}$ and $\tau_{em}$ into Eq. (10) a theoretical estimate of $(1 + z)$ is obtained. It must be borne in mind that $R = (\tau/\tau_{arr})^{2/3}$ and Eq. (6) must be used;

3) the square of the deviation is calculated between the theoretical and the experimental values of $(1+ z)$, multiplied by a weight factor that takes into account the uncertainty of the estimate of $l$;

4) all the quadratic deviations thus obtained are added up. The values of $\lambda$ and $\tau_{arr}$ are changed and one starts again from step 1.

5) The optimal values of $\lambda$ and of $\tau_{arr}$ will be those which minimize the sum of the square deviations. For these values, the correlation index ($r^2$) must be estimated between the theoretical and measured values of redshift, in order to quantify the fit accuracy.

6) By substituting the optimal value of $\tau_{arr}$ into Eq. (8) the value of matter density at present time is obtained.

7) The Hubble parameter in atomic time, estimated at present time, is thus obtained from Eq. (5). The correct value of the "critical" density at present time is then derived from this parameter.

**4.4 Supernova Project: a new data analysis**

The algorithm illustrated in the previous subsection has been applied to the original published data of the Supernova Project [11], relating to 117 type Ia supernovae. The values of the parameters derived from the minimization of the object function were the following:

$$\tau_{arr} = 0.710 \times 10^{18} \text{ s}$$
$$\lambda c^{-2} = 0.637 \times 10^{-56} \text{ cm}^{-2}$$

For each of the 117 supernovae, the theoretical redshift derived from these values was calculated and a representative point was placed on the graph in Fig. 1. The set of points obtained was fitted with a straight line passing through the origin, which proved to be identical with the bisector of the first quadrant (y = 1.0034x); the correlation index was $r^2 = 0.92$.
An estimate of $t_0$ is thus obtained from the relation between $\lambda$ and $t_0$:

$$\lambda = \frac{4}{3 t_0^2} \quad \Rightarrow \quad t_0 = \sqrt{\frac{4}{3 \lambda}} = 4.822 \times 10^{17} \text{ s },$$

i.e. 15.290 billion years.

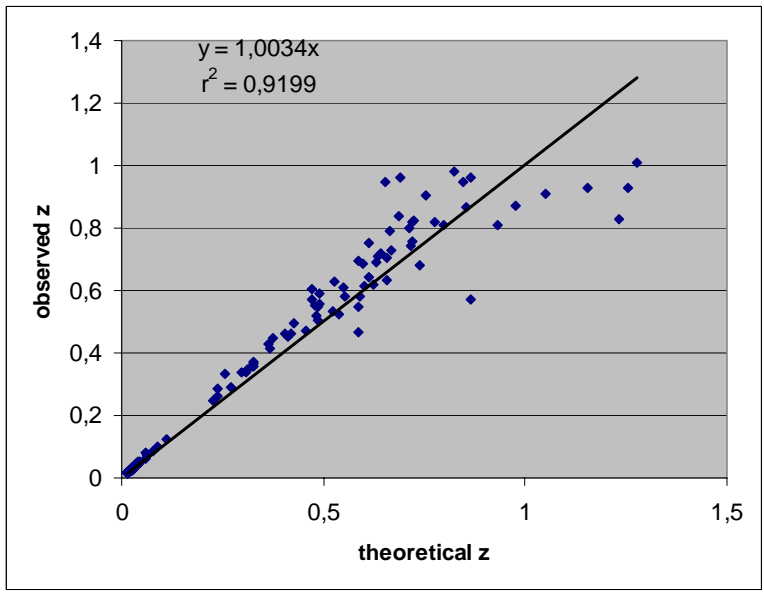

Fig. 1

Linear fit of Supernova Project data

One thus has $\tau_{arr} = (0.710 \times 10^{18}/4.822 \times 10^{17})t_0 = 1.472t_0$. In other words the current time corresponds to a kinematic cosmic time value of $1.472t_0$.

The density of matter at present time is:

$$\mu_0 = \frac{\tau_{arr}^{-2}}{6\pi G} = \frac{1}{6\times 3.14 \times 6.6726\times 10^{-8} \times (0.710)^2 \times 10^{36}} \, \text{g cm}^{-3} = 1.578\times 10^{-30} \, \text{g cm}^{-3}.$$

One thus has:

$$\frac{\lambda}{3} \times \frac{3}{8\pi G} = \frac{\lambda}{8\pi G} = \frac{0.637\times 10^{-56} \times 9\times 10^{20}}{8\times 3.14 \times 6.6726\times 10^{-8}} \, \text{g cm}^{-3} = 3.420\times 10^{-30} \, \text{g cm}^{-3}.$$

The "critical density" is the sum of these two terms and is therefore $4.998 \times 10^{-30}$ g cm$^{-3}$. The contribution of matter is 31%, whilst that of the cosmological term is the remaining 69%.

The value of the Hubble parameter at present time is given by:

$$H_0 = \frac{\dot{R}}{R} = \frac{R'}{R}\frac{d\tau}{d\tau'}\bigg|_{\tau=\tau_{arr}} = \frac{2}{3\tau}\frac{d\tau}{d\tau'}\bigg|_{\tau=\tau_{arr}} = \frac{2}{3\tau_{arr}}\sqrt{1+\frac{3}{4}\lambda\tau_{arr}^2} = 1.67\times 10^{-18}\, \text{s}^{-1} \quad (h=0.52).$$

Rewriting Eq. (7) as:

$$d\tau' = \frac{t_0\, d\left(\frac{\tau}{t_0}\right)}{\sqrt{1+\left(\frac{\tau}{t_0}\right)^2}} \quad \Rightarrow \quad \tau'_{arr} = t_0 \int_0^{\tau_{arr}/t_0} \frac{dx}{\sqrt{1+x^2}}$$

one obtains:

$$\tau'_{arr} = t_0 \ln\left|\frac{\tau_{arr}}{t_0} + \sqrt{1+\left(\frac{\tau_{arr}}{t_0}\right)^2}\right| = 1.179 t_0 = 18\times 10^9 \text{ years}.$$

These results do not essentially alter the already known and agreed framework expressed in the concordance model. Specifically, it is confirmed that the greater part of matter is present in the form of "dark matter". The passage from GR to PGR does not cause dark matter to disappear. One can therefore ask oneself whether it is possible to find an explanation within PGR of the origin of this strange form of matter, similarly to what occurs with "dark energy".

# 5. A POSSIBLE EXPLANATION OF THE "DARK MATTER"

## 5.1 "Cosmic fluid" and the reference acceleration

The Friedmann models are actually based on two hypotheses: general relativity (be it ordinary GR or PGR) and the cosmological principle. The validity of the cosmological principle (e.g., the existence of a scale distance as a function of a "cosmic time") is not a consequence of GR (or of PGR) but a condition imposed on its solutions.

The cosmological principle introduces a substratum of "privileged" inertial frames of reference in respect of which the local physical laws are defined. The existence of this substratum is a fact which is inferred directly from the available experimental data and which can be considered primary. In order to represent this substratum in the context of GR (or PGR) a mathematical fiction is introduced, consisting in the so-called "cosmic fluid". This fluid is characterized by a density $\rho_2$, or by an invariant normalized density $\rho_2 R^3$, with $R$ being the function of scale. This density is the source of a gravitational field; thus the "fluid" is self-gravitating.

The fluid motion is characterized by its local 4-velocity in GR or by its local 5-velocity in PGR. The fluid is defined in such a way that, with respect to the local frame of reference belonging to the inertial substratum, the spatial components of its 4- or 5-velocity vanish. Since, in PGR, the fifth component of the 5-velocity is *locally* null, the only surviving component of fluid velocity (the temporal one) is directed along the time axis of the local frame and points in the direction of the future. The fluid is therefore a fictitious mathematical representation of the real physical fact constituted by the inertial substratum which is implicit in the formulation of the cosmological principle. This fluid is supported by self-gravitation. This requirement imposes specific constraints on the value of $\rho_2$ and on its physical meaning. We now wish to explore, within the PGR approach, the possibility that in general $\rho_2$ <u>is not</u> the true density of matter $\rho_1$. We shall come to consider, at first, the PSR limit. We recall that, according to the PSR redshift-distance relation [1], redshift becomes infinite at distance $r$, so that $r$ can be considered as the radius of the portion of Universe which is visible at a given observer O.

As it is well known [12], the description of a self-gravitating cosmic fluid by means of Euler's and Poisson's equations is equivalent to a description according to which the fluid particle P undergoes gravitational attraction by the fluid mass inside the sphere having its centre in the observer O and radius OP. Moreover, the non-rotating frames of reference having their origin in O and P respectively can be mutually accelerated, even though they are both inertial [12]; in fact, the adjective "inertial" has here to be intended as relative only to the neighborhood of the origin of single frames. Until O and P not compare their measurements of local physical quantities no problem arises.

Let us consider a cosmic fluid particle P of mass $M$, placed on the surface of a homogeneous cosmic fluid sphere having its center in O, radius $x$ and density $\rho_2$. The gravitational potential energy associated with the interaction between the particle and the sphere is easily calculated and results in $-(4/3)\pi G \rho_2 x^2 M$. Let us pose $x = r$; the particle P is then placed at the boundary of the portion of Universe which is accessible to the observer O. The interaction energy is (neglecting the sign) $(4/3)\pi G \rho_2 r^2 M$.

We now note that, in Newtonian terms, the null value of the spatial curvature index $k$ implies that the total energy of cosmic fluid particles is null [12]. Since, by the very definition of cosmic fluid, the four-momentum of its component particles is locally reduced to rest energy, this implies in practice – for the generic fluid particle – the rest energy $Mc^2$ and the gravitational interaction energy with the rest of the fluid to be identical. I.e. it must be:

$$\rho_2 = 3c^2/(4\pi Gr^2) = 3/(4\pi Gt_0^2).$$

This result holds within PSR; when the space expansion is taken into account, it becomes:

$$\rho_2 = 3/(4\pi R^3 Gt_0^2), \tag{11}$$

where the factor $R^3$ has been introduced to take into account the dilution effects resulting from expansion. The scale function $R$ is equal to unity when the radius of the Universe is $r$, and its volume is $(4/3)\pi r^3$.

We can approach the same result from an other point of view. The PSR inertial observer O sees a second inertial observer P, placed at the spatial distance $x$ and at the chronological distance $t$, move at a recession velocity [1, 6] $v = H(t)x$, where $H(t) = H/(1 + t/t_0)$, $-t_0 \leq t \leq 0$, $H = 1/t_0$. The apparent acceleration of P can be obtained by deriving this expression with respect to $-t$. Within the simultaneousness space of O ($t = 0$) this acceleration is:

$$a_{recession} = \frac{c}{t_0}\frac{x}{r} = \frac{x}{t_0^2} \quad ; \tag{12}$$

it is maximum for $x = r = ct_0$ and in such a case it is equal to $c/t_0$; then, according to the point of view of O, the P rest frame of reference is accelerated[2]. The acceleration (12) arises due to the geometry of the contemporaneousness space of O, which is hyperbolic; however, the PGR chronotope is flat (the spatial curvature index $k$ is null). Therefore, in order to convert the PSR substratum into the PGR substratum is necessary the removal of the relative acceleration expressed by Eq. (12).

Let us now introduce a substratum of <u>not</u> mutually accelerated frames and represent this substratum by means of a cosmic fluid, extended up a distance $r$ in every direction around the observer. This means the acceleration of P, expressed by Eq. (12), to be counterbalanced by the gravitational attraction of the fluid mass located between O and P. One has:

$$\frac{4\pi G\rho}{3}x = \frac{x}{t_0^2} \quad \Rightarrow \quad \rho = \frac{3}{4\pi Gt_0^2} = \rho_2, \tag{13}$$

where $\rho$ is the average density of cosmic fluid inside the sphere having its centre in O and radius OP $= x$. As one can see, we face a situation where a local acceleration is equilibrated by a gravitational field in accordance with the GR (or PGR) scheme based on the principle of equivalence.

The PGR real space expansion is easily taken into account by substituting, in Eq. (13), $\rho$ with $\rho R^3$ and $x$ with $(x/R)$, in order to obtain a relation which is independent on cosmic time. We have:

$$\frac{4\pi G}{3}(\rho R^3)\left(\frac{x}{R}\right) = \frac{1}{t_0^2}\left(\frac{x}{R}\right) \quad \Rightarrow \quad \rho = \frac{3}{4\pi Gt_0^2 R^3} = \rho_2 \quad ;$$

that is again the Eq. (11).

In addition, we observe the maximum value of $x/Rt_0^2$ is obtained when $x$ is equal to the radius of the Universe $rR$; it amounts to $c/t_0$. Therefore, the last relation can be written as:

---

[2] Of course, the coordinate translation which moves O into the place P($t$) occuped by P at time $t$ is an inertial transformation. However, the trajectory P($t$) represents an accelerated motion.

$$\frac{4\pi G}{3}(\rho_2 R^3)\left(\frac{x}{R}\right) \leq \frac{c}{t_0} \quad \Rightarrow \quad x \leq rR \ .$$

More generally, posing $\rho = q\rho_2$ one has:

$$\frac{4\pi G}{3}(\rho R^3)\left(\frac{x}{R}\right) \leq \frac{c}{t_0} \quad \Rightarrow \quad x \leq \frac{rR}{q} \ .$$

The maximum $x_{max}$ of $x$ is the distance at which the maximum of the recession acceleration [Eq. (12)] is equilibrated by the self-gravitation. All frames of references having null acceleration relative to O have their origin inside the sphere having centre in the observer O and radius $x_{max}$. This sphere must coincide with the portion of Universe accessible to the observer O; of consequence, it must be $x_{max} = rR \rightarrow q = 1 \rightarrow \rho = \rho_2$.
The possibility $q > 1$ is excluded because, according to it, an observer placed at $x$, where $rR/q < x < rR$, is necessarily falling towards O so that the constraint of null relative acceleration is violated. The possibility $q < 1$ must be discarded as well because an observer placed at $x$, where $rR/q > x > rR$, is constrained with O, while the maximum admissible extension of a constraint is $rR$.

We can clearly recognize a three-step process. The process start with PSR mutually accelerated inertial observers. After the self-gravitational field generated by $\rho_2$ is applied, the recession velocity field $v$ no more depends on $t$, only its dependence on $x$ remains: the relative acceleration of two inertial frames is null. We have passed to the kinematic time scale $\tau$: de Sitter expansion superposes to the real space expansion originated by the appearance of density matter $\rho_2$ [Eq. (5)]. In fact, as it is well known, no static matter distribution undergoing only gravitational forces is possible at the equilibrium. Of course, a material point *travelling* radially respect to the observer O with an apparent velocity $v = v_0 + Hx$ still undergoes an acceleration $dv/dt = H(dx/dt) = Hv = Hv_0 + H^2x$, where $H \approx 1/t_0$. When the proper motion of the point is null ($v_0 = 0$), this acceleration becomes $H^2x$. Finally, after a regraduation $\tau \rightarrow \tau'$ of the time scale (where $\tau'$ is the dynamic time) has been performed, $v$ vanishes and the genuine (not merely projective) space expansion remains. At the same time fictitious repulsive force acting <u>within the flat space</u> appears, which is the cosmological term. This force becomes the cause of the old acceleration $H^2x$, experienced by a material point at rest with respect to O.

Summarizing, we can say the fluid mass density is exactly that required in order to generate, through the self-gravitation, a substratum of not mutually accelerated inertial frames (associated with material bodies) starting from the PSR substratum of mutually accelerated inertial frames (within an empty space). The reasoning set forth shows that the material point P is gravitationally coupled with the density $\rho_2$, not only with real matter. The dynamic action attributed to "dark matter" follows from this result. It is as if the cosmic fluid, or rather its component of density $\rho_2 - \rho_1$ not associable with real matter, were endowed with an actual capacity for gravitational attraction. The situation here reminds to the dynamic effect of "invisible masses" in the context of Hertzian formulation of mechanics (1894) but peculiarly overturned. Whilst Hertz introduced invisible masses to eliminate the forces by means of constraints, a constraint is reproduced here through a (gravitational) force generated by an invisible mass [13].

If the interpretation given is correct, dark matter should be homogeneously distributed in space. Its action, though, should become apparent only around centres of mass of material bodies. Given a sphere having radius $d$ around the centre of mass, a material point placed on the surface of the sphere will interact gravitationally both with the <u>real</u> surrounding mass and with a dark mass

$(4\pi/3)(\rho_2 - \rho_1)d^3$ homogeneously distributed <u>inside</u> the sphere; the densities $\rho_1$, $\rho_2$ must be understood as average densities over entire space[3].

One obtains, from the Ia supernovae data:

$$\rho_2 R^3 = 3/[4 \times 3.14 \times (6.67 \times 10^{-8} \text{ cm}^3 \text{ g}^{-1} \text{ s}^{-2}) \times (0.48 \times 10^{18} \text{ s})^2] =$$

$$= 1.5 \times 10^{-29} \text{ g cm}^{-3}.$$

By dividing this result by the fitted value of the critical density one has:

$$\rho_2 R^3/\rho_{cr} = (1.5 \times 10^{-29} \text{ g cm}^{-3})/(4.998 \times 10^{-30} \text{ g cm}^{-3}) = 3.0 \ .$$

This value is equal to the fitted value of $\Omega_M$ (i.e. 0.31) if a dilution factor equal to $3.0/0.31 = 9.7$ is assumed[4]; this factor will be useful later.

It is a fact worthy of note that with PSR fundamental constants it is possible to define an acceleration that is independent of cosmic time and to which the other accelerations can be compared; this is the acceleration $c/t_0$, counterbalanced by self-gravitation. Its value is approximately $10^{-10}$ m s$^{-2}$, thus it is relevant only at galactic scale or larger. The radial profile of rotation velocity of different spirals presents a maximum corresponding to centripetal acceleration values $\approx c/t_0$. This is the case, for example, of NGC 801, NGC 3198 and NGC 6503.
Hereinafter, we will refer to $c/t_0$ as the "reference" acceleration. The existence of the reference acceleration can explain why there is no reason to doubt Newton's law of gravity when applied to planetary systems, double stars and the like, while the application of the same law to galactic systems makes evident a problem of lack of mass.
It is evident that within the Einstein limit $t_0 \to \infty$ the acceleration $c/t_0$ vanishes. In conventional GR, therefore, there is no possibility of an interpretation of dark matter in the terms represented here. It is important to point out that the dark matter is not a "flaw" of GR or of PGR, but a consequence of the fact that we wish to derive the constraint of null relative acceleration for the fundamental observers by means of a self-gravitating material fluid.

### 5.2 Real matter

Let us now turn to "real" matter and its density $\rho_1$. Let us start by observing that, abstracting from local motions, current lines of the cosmic fluid can be identified as materializations of Universe lines of massive elementary particles. The lightest massive particle created in states of defined mass is the electron. Neutrinos are lighter but are created from weak interactions in oscillating states which are superpositions of eigenstates of the mass.
Let $m$ therefore be the mass of the electron. To travel along the classical electron radius light takes an interval $\theta_0 = e^2/mc^3$. We can thus subdivide the electron Universe line into contiguous intervals of duration $2e^2/mc^3$, each of which corresponds to a possible, distinct creation/localization of the electron. In the archaic metric of the 5-sphere (1), the nucleation phase extends itself from $\underline{x}_0 = 0$ up to $\underline{x}_0 = 2\sigma c\theta_0$, where $\sigma$ is of the order of unity. In the same metric the future of the big bang extends itself from $\underline{x}_0 = 0$ to $\underline{x}_0 = ct_0$. The ratio of the two extensions is therefore $2\sigma\theta_0/t_0 = 2\sigma/N$, where $N = t_0/\theta_0$.

---

[3] Clearly, these assumptions not hold for dark matter strictly associated with single galaxies or clusters. A possible interpretation of this different kind of dark matter is attempted in Sect. 7.
[4] The cube root of 9.7 is 2.13; this is the ratio of scale function values at present time and when the radius of the Universe was $r$. Since the scale function increases by a power of 2/3 of kinematic cosmic time, the ratio of the current cosmic time to the value it assumed then is equal to $(2.13)^{3/2} = 3.11$. Thus the radius of the Universe was $r$ when kinematic cosmic time assumed the value $\tau_0 = 1.472\, t_0/3.11 = 0.47\, t_0$. The same result is derived by inverting Eq. (8). One has: $\mu = 1/6\pi G\tau_0^2 = 3/4\pi Gt_0^2 \to \tau_0 = t_0(2/9)^{1/2} = 0.471\, t_0$.

The *a priori* probability that the creation of the electron - with the materialization of its Universe line - takes place during nucleation is clearly $2\sigma/N$. We can say, therefore, that, on average, the <u>minimum</u> mass required for the creation of a given Universe line at the big bang is $2m\sigma/N$.

Let us now consider the massive creation of matter at the big bang, on the equator of the hypersphere having radius $r = ct_0$ and volume $2\pi^2 r^3$. This space can be subdivided into small cells having volume $(2e^2/mc^2)^3$, each associated with a possible localization of the electron and therefore with a possible Universe line. The number of these small cells is obviously $[(2\pi^2 r^3)/(2e^2/mc^2)^3] = (\pi^2/4)N^3$.

The minimum total mass required for the creation of spacetime is therefore the product of the minimum mass required to create an individual Universe line by the number of distinct Universe lines produced at the big bang, or: $[2\sigma m/N] \times [(\pi^2/4)N^3] = (\sigma\pi^2/2)N^2 m$. We shall assume this to be the mass generated at the big bang.

Let us now consider the instant of cosmic time, subsequent to the big bang, at which the radius of space [in the ordinary metric compatible with Eq. (3)] is $r$ and the volume of space is $4\pi r^3/3$. At this instant the mass density is expressed by $(\sigma\pi^2/2)N^2 m/(4\pi r^3/3)$. If at this instant we assign a value of the scale function $R$ equal to one, the density $\rho_1$ relative to any other instant will be expressed by:

$$\rho_1 R^3 = (\sigma\pi^2/2)N^2 m/(4\pi r^3/3). \tag{14}$$

At this point we obtain an estimate of the true density of matter $\rho_1 R^3$. We have first of all:

$$t_0 = \tau_{arr}/1.472 = (0.710 \times 10^{18} \text{ s})/1.472 = 0.48 \times 10^{18} \text{ s}.$$

$$r = ct_0 = (3 \times 10^{10} \text{ cm s}^{-1})(0.48 \times 10^{18} \text{ s}) = 1.44 \times 10^{28} \text{ cm}.$$

$$e^2/mc^2 = 2.82 \times 10^{-13} \text{ cm}; \quad m = 9.11 \times 10^{-28} \text{ g}.$$

Thus:

$$N = [(ct_0)/(e^2/mc^2)] = (1.44 \times 10^{28} \text{ cm})/(2.82 \times 10^{-13} \text{ cm}) = 0.51 \times 10^{41}.$$

$$4\pi r^3/3 = 1.25 \times 10^{85} \text{ cm}^3.$$

$$\rho_1 R^3 = \sigma(12.8 \times 10^{81})(9.11 \times 10^{-28} \text{ g})/(1.25 \times 10^{85} \text{ cm}^3) = 9.3\sigma \times 10^{-31} \text{ g cm}^{-3}.$$

To obtain the value of the density corresponding to the present time we ought to divide this result by the cube of the ratio of scale function values at present time and when the volume of space was $4\pi r^3/3$. The value of this dilution factor was obtained in the previous subsection and is equal to 9.7; one has:

$$\rho_1 = 0.96\sigma \times 10^{-31} \text{ g cm}^{-3}.$$

By dividing this value by the value of the critical density at present time one obtains:

$$\Omega_{Mreal} = (0.96\sigma \times 10^{-31} \text{ g cm}^{-3})/(4.998 \times 10^{-30} \text{ g cm}^{-3}) = 0.019\sigma.$$

Assuming $\sigma \approx 3.9$ and bearing in mind that $h = 0.52$, this result agrees both with the estimates of baryonic matter density derived from Boomerang and COBE/DMR and with those based on primordial deuterium measurements in quasars. It suggests a real mass that is practically entirely concentrated in baryons.
The ratio of the $\rho_1 R^3$ and $\rho_2 R^3$ densities is clearly independent on cosmic time and is equal to:

$$\alpha = (9.3\sigma \times 10^{-31} \text{ g cm}^{-3})/(1.5 \times 10^{-29} \text{ g cm}^{-3}) = 0.242 \ .$$

In other words, cosmic fluid is approximately four times denser than real matter.

## 6. TO WHAT EXTENT IS A FRAME OF REFERENCE INERTIAL? PIONEER ANOMALY AND EARTH FLYBYS

The problem has often been discussed in the literature of anomalous accelerations undergone by the Pioneer probes after flybys of external planets or by several spacecraft during Earth flybys [14, 15]. Specifically, the anomalous acceleration of the Pioneer probes is of the same order of magnitude as $c/t_0$, and this has led to diverse speculations on the origin of this anomaly somehow connected with the subjects treated here. In our view, various controversial aspects must be cleared up before this connection can be stated with certainty. Specifically, an interesting detail is that both in the case of the Pioneer probes and in that of terrestrial satellites, the anomalous acceleration detected is of the same order of magnitude as the non-inertiality of the most commonly used frames of reference. This relevant detail, in our best knowledge, is not addressed in current literature.

To illustrate this topic, let us consider a body in uniform circular motion on a circumference of radius $R$ with a revolution period $T$. The radial acceleration (the only one in existence) of this body is therefore:

$$a = 4\pi^2 R/T^2 \ . \tag{15}$$

By first approximation, the frame of reference in rest respect to the fixed stars having as its origin the centre of mass of the Earth can be deemed an inertial frame because, by substituting into Eq. (15) the radius of Earth's orbit ($1.5 \times 10^{11}$ m) and Earth's revolution period ($3.1 \times 10^7$ s), an acceleration resulting from Earth's revolution is found that is only

$$a_{\text{rev}} = (59.21 \times 10^{11} \text{ m})/(9.61 \times 10^{14} \text{ s}^2) = 6.2 \text{ mm s}^{-2} \ .$$

It is therefore natural to refer the motion of satellites to this frame (frames implementing the ITRS system). It is clear, however, that when satellite accelerations are measured with an accuracy of mm s$^{-2}$ the perturbation due to the non-inertial nature of this frame is no more negligible. In this case, the satellite motion should be referred to a more inertial frame, for example that which is in rest respect to the fixed stars and has as its origin the Sun centre of mass (frames implementing the ICRS system).

In any case, even this frame of reference is not exactly inertial. By substituting into Eq. (15) the radius of the Sun's orbit around the centre of the galaxy ($0.3 \times 10^{21}$ m) and the period of the Sun's revolution around this centre ($7.73 \times 10^{15}$ s), respectively, one obtains an acceleration:

$$a_{\text{Sun}} = (11.84 \times 10^{21} \text{ m})/(59.75 \times 10^{30} \text{ s}^2) = 2 \times 10^{-10} \text{ m s}^{-2} \ .$$

This acceleration value is much less than the previous one, and this frame is therefore appropriate for studying satellite motion in the circumstances mentioned previously. It is evident, however, that if one measures the acceleration of a space vehicle with a degree of accuracy that is no longer of mm s$^{-2}$, but of $1 \times 10^{-10}$ m s$^{-2}$ or less, the non-inertiality of this second system also is no more negligible. The motion of the vehicle will then have to be referred to an even more approximately inertial frame, for example a frame having its origin in the galactic centre. It is at this final level that the effects linked to the existence of some reference acceleration should become apparent.

The anomalous effects mentioned with regard to satellites correspond to unexplainable accelerations of the order of a few mm s$^{-2}$. Is it possible that their motion was referred to the Earth, and that part of the anomaly merely derived from the imperfect inertiality of the Earth frame?
The Pioneer effect corresponds precisely to anomalous accelerations of the order of $10^{-10}$ m s$^{-2}$. Such accelerations, oriented towards the Sun, were deduced from Doppler measurements of the radial velocity of the probes. Is this a possible result of the imperfect inertiality of the frame adopted?

It is difficult, therefore, to separate any effects associated with some reference acceleration from the other systematic factors in play, which are of the same or of a greater order of magnitude. All this, together with a large number of other sources of uncertainty (dynamic forces resulting from radiation pressure, loss of gas from the probe, etc.) makes it, in our opinion, highly difficult to use the data on anomalies for reliable speculation on the nature of inertia or on dark matter.
Space missions specifically designed for the purpose, with accurate accelerometry in the three dimensions of space and complete kinematic determinations, would probably be useful.

## 7. DARK MATTER ASSOCIATED WITH GALAXIES AND CLUSTERS

The dark matter specifically associated with a single galaxy or cluster represents a phoenomenon which is in principle different by the global "cosmological" dark matter whose nature has been discussed before.
Let us consider the substratum of not mutually accelerated fundamental observers defined before. The simplest assumption is these observers are inertial (that is, according to them the local motion of a free test particle should be rectilinear and uniform) but this could not be the case. For example, is possible to speculate about the existence of a small acceleration of a free test particle with respect to them. We can conjecture this acceleration $a'$ is:

$$a'(\vec{x}) = \frac{c}{t_0} \sum_i \begin{cases} \left(\frac{|\vec{x}-\vec{x}_i|}{d_i}\right) \frac{\vec{x}_i - \vec{x}}{|\vec{x}-\vec{x}_i|} & \text{for } |\vec{x}-\vec{x}_i| \leq d_i \\ 0 & \text{otherwise} \end{cases} \qquad (16)$$

where $\vec{x}$ is the particle position and the random variables $\vec{x}_i$ are uniformly distributed on the space of cosmic time $\tau$ = constant. If it is so, only the (non-rotating) fundamental observers placed in points where $a' = 0$ will be really inertial; the others will be only approximatively inertial. When $a' \rightarrow 0$, an inertial substratum is reobtained which coincides with the entire *continuum* of not mutually accelerated observers. In other words, the non-rotating frame of reference whose origin is the material point P considered in Sect. 5.1 is inertial provided that $a'(P) = 0$, otherwise it is not. When the self-gravitation is applied and the PSR escape acceleration [Eq. (12)] is counterbalanced by it, this frame is no more accelerated with respect to an inertial observer O; yet, this fact not influences $a'(P)$.

According to Eq. (16), the cosmical matter is forced to condensate around the points $\vec{x}_i$, which play the role of condensation nuclei, so that bounded states of dimension $d_i$ are generated. Due to the uniformity of their spatial distribution, the linear density of these nuclei along a given direction not presents any directional dependence, as requested by the cosmological principle. In addition, their relative distances not vary, except for the space cosmical expansion[5].

Within a not accelerated frame of reference having its origin in proximity of one of these nuclei, the inertia principle is violated; a fictitious force appears, giving to physical bodies an acceleration towards the nucleus whose maximum equates the reference acceleration. This hypothesis may be relevant in reference to the dynamics of galaxies and clusters; in particular, a more or less severe freezing of peculiar motions at different scales arises.

Incidentally, we remark the acceleration $a'$ could be the true cause of several effects commonly attributed to the dark matter, for example anomalies in radial velocity profiles of spirals. A material point in equilibrium, placed at the border of a spiral galaxy, must revolve with a radial acceleration equal and opposite to $a' \approx c/t_0$, as if a supplementary mass $M'$ were attracting it. The ratio $M'/M$, where $M$ is the real mass, is of the same order than $(c/t_0)/(v^2/d)$, where $v$ is the velocity of the point in the limit case $a' = 0$ and $d$ denotes its distance from the galactic centre. For a spiral typical, this ratio falls in the interval 0.1-100 so revealing the importance of $a'$ in galactic dynamics.

More generally, let us consider a bounded system of dimension $d$, whose centre of mass coincides with a condensation nucleus and let us denote with $y$ the distance from this nucleus. If it so, Eq. (16) is approximated as:

$$a' = -\frac{c}{t_0}\frac{y}{d} . \qquad (17)$$

This is the same acceleration $\omega^2 y$ of a harmonic oscillator of frequency $v$, where:

$$v = \frac{\omega}{2\pi} = \frac{1}{2\pi}\sqrt{\frac{c}{dt_0}} . \qquad (18)$$

The same result is obtained if, at the system border $y \approx d$, the acceleration (17) is equilibrated by a centrifugal force $v^2/d$, where $v = \omega d$. As an application, substituting $d = 10^{22}$ m, approximatively the mean distance of two spirals within a cluster, in Eq. (18) we obtain $T = 1/v = 25 \times 10^{15}$ s, to be compared with the period of Sun revolution around the galactic centre, $7.72 \times 10^{15}$ s.

## 8. GEOMETRY OF PRIVATE SPACETIME

This section is devoted to the calculation of the "private" metric coefficients $g_{ij}$ provided by the model, expressed in physical coordinates directly connected with the observations. In particular, we aim to evidence if any effect related to the small projective deviations from the usual RW metric can be detected at the scale of the Solar System with present day observational methods.

The projective metric can be written as:

$$ds^2 = d\bar{x}_0 d\bar{x}^0 - R^2(\tau)\left[d\bar{x}_1 d\bar{x}^1 + d\bar{x}_2 d\bar{x}^2 + d\bar{x}_3 d\bar{x}^3\right] + d\bar{x}_5 d\bar{x}^5 . \qquad (19)$$

---

[5] Therefore, we can suppose these nuclei are generated as a relic of inertia fluctuation within the archaic vacuum. A paper discussing this subject is currently in preparation.

The relation between projective and physical coordinates is given by:

$$\bar{x}_\mu = x_\mu / A; \quad \mu = 0,1,2,3; \qquad \bar{x}_5 = r/A \ . \tag{20}$$

where:

$$A^2 = 1 + \frac{x_0 x^0}{r^2} - R^2(\tau)\frac{x_i x^i}{r^2}; \quad i = 1,2,3. \tag{21}$$

Moreover [3]:

$$\bar{x}_0 = r\frac{\dfrac{\tau}{t_0}}{\sqrt{1 + \left(\dfrac{\tau}{t_0}\right)^2}} \ . \tag{22}$$

The relation between atomic and kinematic time can be re-written as:

$$d\tau' = \frac{d\tau}{\sqrt{1 + \left(\dfrac{\tau}{t_0}\right)^2}} \ . \tag{23}$$

Therefore:

$$d\bar{x}_0 = \frac{c\, d\tau'}{1 + \left(\dfrac{\tau}{t_0}\right)^2} \ ; \tag{24}$$

$$ds'^2 = \frac{c^2 d\tau'^2}{\left[1 + \left(\dfrac{\tau}{t_0}\right)^2\right]} - R^2(\tau')\left[d\bar{x}_1\, d\bar{x}^1 + d\bar{x}_2\, d\bar{x}^2 + d\bar{x}_3\, d\bar{x}^3\right] + d\bar{x}_5\, d\bar{x}^5 \tag{25}$$

where:

$$ds'^2 = ds^2 \frac{d\tau'^2}{d\tau^2} = \frac{ds^2}{\left[1 + \left(\dfrac{\tau}{t_0}\right)^2\right]} \ ,$$

because we are passed from the $\tau$-clock to the $\tau'$-clock, so that the rhythm of proper time is changed. Substituting Eqs. (20) and (22) in Eq. (21) one obtains:

$$A^2 = \left(1 - R^2 \frac{l^2}{r^2}\right)\left[1 + \left(\frac{\tau}{t_0}\right)^2\right], \tag{26}$$

where $l = (x_i x^i)^{1/2}$ is the space distance from the observer.
Differentiating Eqs. (20) one has:

$$d\bar{x}_5 = -r\frac{dA}{A^2}; \quad d\bar{x}_i = \frac{dx_i}{A} + \frac{x_i}{r}d\bar{x}_5 \ ; \quad i = 1,2,3. \tag{27}$$

Differentiating Eq. (26) one has:

$$dA = \frac{d(A^2)}{2A} = \frac{x_0 dx^0}{r^2 A} - R^2\frac{x_i dx^i}{r^2 A} = \frac{\bar{x}_0 dx^0}{r^2} - R^2\frac{x_i dx^i}{r^2 A} =$$

$$= \frac{\left(\frac{\tau}{t_0}\right)}{r\sqrt{1 + \left(\frac{\tau}{t_0}\right)^2}} \frac{dx_0}{d\tau} d\tau - R^2\frac{x_i dx^i}{r^2 A} \ .$$

We remark that:

$$\frac{d\tau}{dx_0} = \frac{d\tau}{d\bar{x}_0}\frac{d\bar{x}_0}{dx_0} = \frac{1}{c}\left[1 + \left(\frac{\tau}{t_0}\right)^2\right]^{3/2} \frac{d}{dx_0}\left(\frac{x_0}{A}\right) = \frac{1}{cA}\left[1 + \left(\frac{\tau}{t_0}\right)^2\right]^{3/2}\left(1 - \frac{x_0^2}{r^2 A^2}\right).$$

From Eqs. (20), (22) the following relation is derived:

$$\frac{d\tau}{dx_0} = \frac{1}{cA}\sqrt{1 + \left(\frac{\tau}{t_0}\right)^2} \quad \Rightarrow \quad \frac{dx_0}{d\tau} = \frac{cA}{\sqrt{1 + \left(\frac{\tau}{t_0}\right)^2}}.$$

Therefore:

$$dA = \frac{cd\tau'}{r}\frac{A\left(\frac{\tau}{t_0}\right)}{\sqrt{1 + \left(\frac{\tau}{t_0}\right)^2}} - R^2\frac{x_i dx^i}{r^2 A} \ . \tag{28}$$

Substituting Eqs. (27) and (28) in Eq. (25) and rearranging the terms one finally obtains:

$$ds^2 = g_{00} d\tau'^2 + 2g_{i0} d\tau' dx^i + g_{ii} dx^i dx^i; \quad i = 1,2,3. \tag{29}$$

This metric is expressed in physical coordinates and the usual atomic time is used[6]. One has:

$$g_{00} = \frac{1}{1+\left(\frac{\tau}{t_0}\right)^2} + \left[1+\left(\frac{\tau}{t_0}\right)^2\right]\left(1-R^2\frac{l^2}{r^2}\right)^2\left(\frac{\tau}{t_0}\right)^2\frac{1}{A^4};$$

and recalling Eq. (26):

$$\left[1+\left(\frac{\tau}{t_0}\right)^2\right]g_{00} = 1 + \frac{1}{A^4}A^4\left(\frac{\tau}{t_0}\right)^2 \quad \Rightarrow \quad g_{00} = 1. \tag{30}$$

Analogously:

$$g_{0i} = \frac{2R^2 x_i}{rA^2}\frac{\left(\frac{\tau}{t_0}\right)^3}{\sqrt{1+\left(\frac{\tau}{t_0}\right)^2}}. \tag{31}$$

$$g_{ij} = -\frac{R^2 \delta_{ij}}{\left(1-R^2\frac{l^2}{r^2}\right)} - \frac{R^4 x_i x_j}{r^2 A^4}\left[1+2\left(\frac{\tau}{t_0}\right)^2\right]. \tag{32}$$

How one can see, in the Einstein limit $r, t_0 \to \infty$ as well as in the local limit $l, x_i \to 0$ (that is, in the proximity of the observer) the off-diagonal coefficients vanish, while the diagonal coefficients collapse on the ordinary RW metric.
Within the Solar System ($l, x_i \approx 10^{13}$ m) $g_{0i}$ coefficients are of the order $10^{-13}$, while $g_{ij}$ off-diagonal coefficients are of the order $10^{-26}$. Under the same circumstance, the deviation of diagonal coefficients from RW is also of the order $10^{-26}$. Considering that more recent local tests of determination of the metric coefficients present accuracies of the order $10^{-4}$, $10^{-5}$, it seems reasonable conclude that no projective effect on $g_{ij}$ can be detected at present by means of local tests.

---

[6] We remark that $g_{0i}$ coefficients are simmetric, contrary to the claim of Ref. [3]. This is due to an algebraic error, here emended.

## 9. CONCLUSIONS

The main difference between usual Friedmann cosmology and the PGR approach consists in the introduction of a new fundamental constant of nature, identical for all observers at any cosmic time, which is $t_0$. Several effects of the finite value of this constant have been presented; these effects disappear in the limit $t_0 \to \infty$ which corresponds to ordinary General Relativity. Specifically:

1) The density of matter measured in accordance with the public metric valid "before" nucleation is finite, because the volume of the equator of the Arcidiacono hypersphere is finite.
2) The density of matter measured in accordance with the private metric of the individual fundamental observer after nucleation is finite because the space contraction factor with respect to the public metric is finite.
3) Also the difference between commonly used atomic cosmic time and kinematic cosmic time is a result of the finite value of $t_0$. Such a difference implies a slight re-evaluation of cosmological parameters which does not alter the currently agreed cosmological framework, as emerges from a new analysis of Supernova Project data. The appearance of the cosmological term (dark energy) is an effect of this difference.
4) The purely kinematic expansion also detected in the limit case of null density (PSR) causes a universal constant to appear having the size of an acceleration, $c/t_0$. This constant is clearly null in the limit of ordinary GR.
5) This "reference" acceleration appears in different situations: from the maximum of radial velocity profile of spirals to the Pioneer effect. The experimental situation, however, requires some clarification.
6) Physical laws are defined with reference to a substratum of not mutually accelerated, inertial observers.
7) If this substratum is described as a self-gravitating fluid, the self-gravitation must then balance the escape acceleration whose maximum value is equal to the reference acceleration. The average density of this fluid must therefore be equal to the critical density.
8) The greater part of the fluid density, however, does not correspond to real matter, because it actually describes the constraint imposed under point 7). Dark matter thus originates.
9) If one assumes that the density of real matter is just sufficient to create a spacetime structure, it can be estimated and is seen to be finite because the value of $t_0$ is finite. The abovesaid density of real matter can be made to agree with primordial baryogenesis and one can therefore suppose that the greater part of real mass is concentrated in baryons.
10) The projective deviations of PGR metric coefficients from usual RW does not seem to have any observational consequence detectable with present day technology.

PGR therefore seems to allow many troublesome difficulties to be overcome which affect the standard version of the big bang (the appearance of singularities, the problem of the origin of dark matter and of dark energy, the flatness enigma, etc.). At the same time, it opens up interesting scenarios for future research connected with the experimental and observational highlighting of reference acceleration in its relation with "dark matter". Also the possible relation with other theoretical attempts as, for example, $F(R)$ gravity (16,17) remains to be investigated. We can therefore conclude by saying that the paradigm proposed many years ago (18,19,20) by Fantappié and Arcidiacono is as topical as ever and deserving of further study.

ACKNOWLEDGMENTS

I wish thank my friend Ignazio Licata for many discussions and constant encouragement. I'm in debt with Prof. Corda for useful advice and support.